# Collaborative Brain-Computer Interface for Human Interest Detection in Complex and Dynamic Settings

Amelia J. Solon
DCS Corporation
Army Research Laboratory
Aberdeeen Proving Ground, MD, USA
asolon@dcscorp.com

Stephen M. Gordon
DCS Corporation
Army Research Laboratory
Aberdeeen Proving Ground, MD, USA
sgordon@dcscorp.com

Jonathan R. McDaniel
DCS Corporation
Army Research Laboratory
Alexandria, VA, USA
jmcdaniel@dcscorp.com

Vernon J. Lawhern
Army Research Laboratory
Aberdeeen Proving Ground, MD, USA
vernon.j.lawhern.civ@mail.mil

*Abstract*— Humans can fluidly adapt their interest in complex environments in ways that machines cannot. Here, we lay the groundwork for a real-world system that passively monitors and merges neural correlates of visual interest across team members via Collaborative Brain Computer Interface (cBCI). When group interest is detected and co-registered in time and space, it can be used to model the task relevance of items in a dynamic, natural environment. Previous work in cBCIs focuses on static stimuli, stimulus- or response- locked analyses, and often within-subject and experiment model training. The contributions of this work are twofold. First, we test the utility of cBCI on a scenario that more closely resembles natural conditions, where subjects visually scanned a video for target items in a virtual environment. Second, we use an experiment-agnostic deep learning model to account for the real-world use case where no training set exists that exactly matches the end-users' task and circumstances. With our approach we show improved performance as the number of subjects in the cBCI ensemble grows, and the potential to reconstruct ground-truth target occurrence in an otherwise noisy and complex environment.

## I. INTRODUCTION

The human brain is a highly flexible and adaptable pattern recognition system. Our brains enable us to accurately perceive the world in the face of changing environments and task demands. This flexibility cannot yet be replicated by current artificial intelligence systems. In this work, we propose a system which models a team of humans as a set of distributed sensors in a naturalistic and dynamic environment. If the team shares a common goal, the system can leverage that commonality to model spatial and temporal characteristics of task relevant items in the environment, which can then be shared with the individual teammates or with other, possibly autonomous, agents.

Directly querying an individual [1] for their assessment of a scene, or synthesizing of information across group members [2, 3], may be deleterious to their primary tasking in complex scenarios. As such, our proposed system does not query or require in-the-loop feedback from the user and, thus, does not interrupt the tasking of any teammate. It instead passively monitors, in real-time, each teammates's electroencephalogram (EEG) signals and eye movements via Brain Computer Interface (BCI) technology.

BCIs, classically researched as medical devices [4] and recently explored to enhance the capabilities of healthy users [5], are equipped with decoding algorithms that translate neural activity into communication and control signals. The BCI models in this system are trained, from EEG signals, to detect whether or not an individual has seen an object of interest. There is a wealth of prior research on EEG potentials coincident with task relevant, attention grabbing, or visually salient stimuli [6]. Among them is the P300 response, a positive voltage potential occurring over the visual cortex about 300ms after a target stimulus, which is thought to play a critical role in gating low-level perception to higher-order memory [7]. Because the P300 response sits between initial perception and cognitive function, it is the focus of many BCI designs in the context of perception and decision-making.

The proposed system merges BCI scores (here, the probability that a trial of EEG contains a response to an interesting visual stimulus) across individual teammates experiencing the same stimulus, into a Collaborative Brain Computer Interface (cBCI) framework. The ensemble approach inherent in cBCIs has the potential to mitigate noise and error on both the individual user's and BCI algorithm's part. Individual perception is often inferior to group perception [8], and single-trial detection of P300 signals remains notoriously challenging in complex environments. Indeed, prior work shows that cBCIs generate higher performance compared to individual BCIs across a wide variety of BCI paradigms [9-16]. We believe that a cBCI approach is the best starting point to ensure that only those items with the highest probability of task relevance are added to the overall environmental representation.

We envision this system applied to a team of experts operating in a complex, natural, and potentially ambulatory environment. While cBCI designs continue to surpass the performance of individual BCIs in increasingly complex visual tasks [9-12], previous work generally trains BCI models for specific applications that are stimulus- or response- locked, and leverage static stimuli. Due to the unpredictable nature of a real-world environment, BCIs may not have access to stimulus presentation information or user responses, and visual targets often enter the field of vision (FOV) dynamically. Items can suddenly appear in the FOV, whereas others may only elicit a response when fixated upon. In addition to exogenous variables, a human's endogenous state can vary in ways that affect the nature of the evoked response. Therefore, cBCI models applied to real-world scenarios must be able to 1) operate in free-viewing visual search tasks with naturalistically occurring targets, and 2) generalize to new domains since training data matching the exact, unpredictable, real-world use-case cannot be replicated in the lab.

Here, we specifically address these two issues by using an experiment-agnostic deep learning model, and investigating the applicability of a cBCI approach in an unconstrained visual search (i.e. free-viewing) task. In previous work, we showed that a deep learning model trained on a pool of multiple target-detection experiments (with different event-locking and cognitive state variables) generalized better to unseen scenarios. Finally, as a proof-of-concept for natural, free-viewing tasks, we test, offline, on a dataset where subjects visually scanned a dynamic environment (video) for threats in both high and low visibility conditions.

## II. BACKGROUND

### A. Collaborative Brain Computer Interfaces

The goal of the proposed system is to identify task relevant objects in a dynamic environment. However, each individual will perceive relevant and distracting stimuli differently. Therefore, if a BCI model is 100% accurate for an individual, this does not mean that it is 100% accurate for the task. In this case, collaboration across people has the potential to detect only the stimuli that are related to the groups' shared goals.

Originally, individual event-related-potential (ERP) BCIs averaged EEG responses to multiple instances of the same stimulus to achieve high classification accuracies [17]. To avoid the time delays imposed by averaging multiple trials from a single user, researchers have explored averaging single trials collected from multiple users experiencing the same stimuli [12]. Today, individual BCI designs rarely use averaged ERP approaches, instead opting for single-trial classification of the EEG data [18,19]. In these cases, cBCI framework shows improved classification compared to their individual BCI counterparts, even across a wide array of paradigms [9-16]. This could be due to several factors. Through the same ensemble processes that can improve signal-to-noise ratio (SNR) or classifier stability, cBCIs provide a means for measuring group perception [8]. cBCIs have the potential to detect group interest without the need for standard communication between group members. This is especially useful in scenarios where communication between teammates could hurt performance outcomes [2,3].

Previous work on ERP-based cBCI systems generally focuses on improving accuracy and speed of the system over that attained by group behavior. The speed and accuracy gains often come at a price (the efforts of multiple BCI users) for simple tasks that may not scale to real-world applications. However, several cBCI works have made important steps towards real-world use. In [9], subjects were shown complex natural images of an arctic environment, where they were tasked with identifying polar bears amidst a crowd of penguins, building upon previous work where subjects were instead shown artificial images [20]. In [10], the same researchers successfully transferred models from [20] to the task in [9] to bypass within task training. Both experiments used the same stimulus presentation paradigm: static images were presented in set sequence that allowed participants to return to baseline from the previous trial. The cBCI leveraged neural and behavioral features, and weighted each user's contributions based on features related to the confidence in their decision. In [16], cBCI improved performance over individual BCI for auditory P300 detection. These systems were additionally equipped with Global Positioning Systems to facilitate ambulatory life logging. In this work we to take steps to develop a visual interest detection cBCIs for use in non-stimulus locked and dynamic natural environments.

### B. Pooled-Experiment Visual Target Detection Models

The previously discussed cBCI approaches utilize BCI models trained per-user and per-application (with the exception of [10]). While these specific classifiers yield higher performance, they are contingent on training data that match the exact user and end application. This assumption is valid for laboratory experiments, games, and several assistive technologies, but our previous research indicates that this may be problematic for real-world applications [23,24].

In juxtaposition to many laboratory experiments, the real world is often not delimited into clear instances of target, or task-relevant stimuli, and "everything else". The relevance of a particular stimulus to a given task is a function of both the task and the human performing the task, i.e. different people will place varying levels of importance on different objects. Classic P300 studies, as well as P300-based BCIs, assume that there are at most three types of stimuli: target, distractor, and background. The goal is to separate the targets from both the distractors and background. Because distractors elicit attenuated P300 responses [21], BCI designers optimize performance by fitting models to the specific target stimuli (i.e. task) and to the user. Yet, if the task changes or if distractors are important [22], these systems can quickly become suboptimal.

We previously presented work in which we used deep learning methods to construct across-experiment BCI systems [23]. In other words, we trained a BCI model using data from one experiment and set of subjects and applied that model to another, unseen, experiment and set of subjects. As expected, these across-experiments models performed worse when compared to within experiment models [24]. However, when we pooled together multiple experiments, thus increasing the amount and diversity of training data, the average performance on unseen test sets increased. The net effect being that the pooled-experiment models provided the best performance if the exact task or stimuli was unknown or could change [23]. We believe that these pooled-experiment models are a necessary component for translating BCI, collaborative or otherwise, into real-world applications.

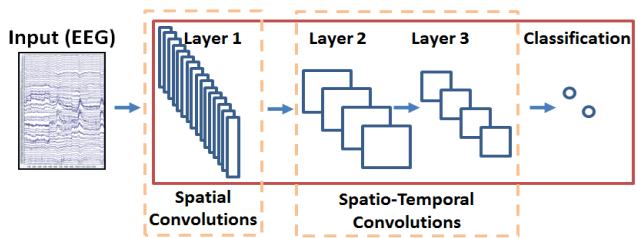

Fig. 1. EEGNet architecture.

## III. METHODS

### A. BCI Model Development

#### 1) Model Architecture

For our BCI model, we use the EEGNet Deep Learning architecture [25]. The architecture is inspired by standard temporal and spatial filters often used in EEG feature extraction. Previously, we showed that EEGNet enabled cross-subject transfer performance equal to or better than conventional approaches for several BCI paradigms. EEGNet is also the model we used to obtain our cross-experiment results described in [23,24].

Fig. 1 shows the general architecture of EEGNet. Given an input trial of minimally pre-processed time-series EEG data, with $C$ channels and $T$ time points, we learn 16 spatial filters, implemented as a convolution across the channel dimension (Layer 1), to reduce dimensionality and improve SNR. Layers 2 and 3 each use four spatio-temporal convolutions to learn correlations in time and across spatial filters. The classification is a two unit dense layer with a softmax activation function. The model was trained with the binary cross entropy loss function, and used the ADAM optimizer. All convolutional layers used batch normalization, Dropout and $L_2$ regularization to mitigate overfitting. For more details, the interested reader is referred to [25].

#### 2) Model training

We train an EEGNet model on a pool of multiple experiments' and subjects' data, described in Table I. Although each dataset has a unique experimental design, they each have a visual target detection task. Subjects either mentally counted or pressed a button in response to each target occurrence. Analyses were either stimulus-locked to rapid serial visual presentation (RSVP) of images, or fixation-locked to guided fixations. Some experiments only contained targets (T) and background (B), while others also contained distractors (D). When present, distractors were included in the target class, as they elicit attenuated P300 responses [21].

With the exception of the 5Hz RSVP dataset, all experiments in Table I were recorded with a 64-channel BioSemi Active II EEG System. The 5Hz RSVP data were recorded using a 256 channel BioSemi Active II; channels were spatially downsampled to match the 64 channel montage of the other datasets. All experiments were approved by the Institutional Review Board of the Army Research Laboratory.

TABLE I. LIST OF TRAINING DATA

| Time Locked To: | # of Subjects | Total # Instances (T/D vs B)* | Response to Target Stimuli | Experiment Description |
|---|---|---|---|---|
| Stimulus | 18 | 12,965 291,854 | Button | 5Hz RSVP with varying target difficulty [26] |
| Fixation | 16 | 2,658 28,030 | Button | Guided fixations with variable workload [27] |
| Stimulus | 16 | 10,512 99,504 | Count or Button | 2Hz RSVP with static or moving targets [28] |
| Stimulus | 10 | 269 998 | Count | 1Hz RSVP; free choice target detection |
| Stimulus | 20 | 5,401 37,799 | Button | 1Hz RSVP before / after physical exertion [29] |
| TOTAL: | 80 | 31,805 458,185 | --- | --- |

a. Target (T), Distractor (D), and Background (B) stimuli

All data were bandpass filtered between 0.3 Hz and 50 Hz before being downsampled to a 128 Hz sampling rate. If the dataset was stimulus-locked or had guided fixations, epochs were extracted [0s 1.25s] around stimulus presentation.

We balanced the training set by randomly under-sampling the majority class (non-targets) for each subject within an experiment and limit the number of balanced training instances per experiment to 6000 (about 3000 for each class). We train five EEGNet models, each balanced with different random selections of the full training data collection, to ensure better coverage of the total available training data. We ensemble the five models by averaging the classifier scores per test instance.

### B. Test Set: Free-Viewing Target Detection in Video

Our hold out test dataset is a Free-Viewing (FV) task in which participants (16 male, avg. age 28.3) viewed an urban landscape in a 15 minute video. Rather than view a set of static images, participants were "driven" through a virtual environment (Fig. 2) and asked to look for two different types of items and discriminate between visually similar threats (a man with a weapon, or table oriented such that it could hide an explosive device) and non-threats (a man without a weapon, or a table that one could see under). Targets would abruptly appear one at a time in random but logical locations (i.e. on the street or in a doorway) at an approximate rate of once every three seconds. Targets stayed on screen for one second before disappearing [24]. Participants were free to scan the environment but were instructed to indicate the type of target they observed by pressing a button with either the left or right index finger. Participant responses were graded for speed and accuracy and a score was given for each response. The cumulative score was displayed at the top and bottom of the screen. At different times in the video, a dense fog was overlaid on the scene.

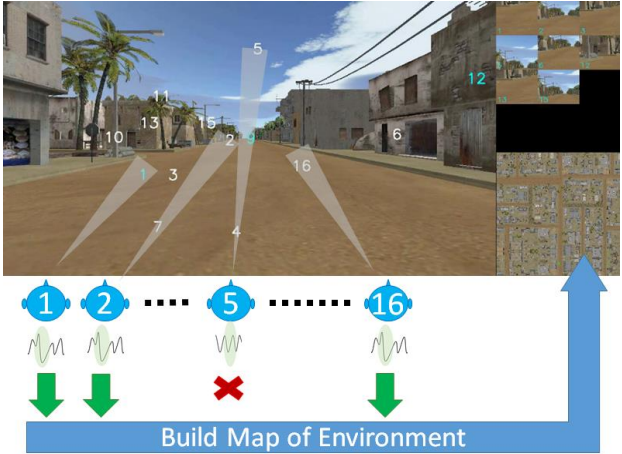

Fig. 2. Depiction of test dataset and proposed system. Numbers on the screen indicate the gaze position for the corresponding participant. For example, Participant 1's gaze falls where the number '1' is located on screen.

Targets "pop up" on the screen in both Fog (low visibility) and No-Fog (high visibility) conditions. Such precise onsets do not represent the majority of visual stimuli that a person experiences day-to-day. While the evoked responses are still bounded by target item onset and offset in the Fog condition, we expect the attention-grabbing pop-up effect to be attenuated due to the obfuscation of targets by the fog overlay. Since we expect that subjects will have to deliberately scan the scene to find targets, we believe that the Fog condition more closely resembles real-world conditions. Therefore the No-Fog and Fog conditions will be analyzed separately.

Horizontal and vertical electrooculogram (EOG) data were recorded, respectively, by placing electrodes near the outer canthus of each eye, and at the pupil. EEG data were recorded with a 64-channel BioSemi Active II EEG System. The experiment was approved by the Institutional Review Board of the Army Research Laboratory. Data were bandpass filtered between 0.3 Hz and 50 Hz before being downsampled to a 128 Hz sampling rate.

The discrimination task for the EEGNet models was to label target (both threat and non-threat items) versus search fixations. In free-viewing experiments, processing of the stimuli can begin before the fixation is complete, showing earlier P300s than in visual oddball tasks [30]. As such, data were epoched [-0.3s, 0.95s] around all fixation events, which were identified via EOG signals using a per-subject EOG velocity threshold. Even though items appeared to subjects at the same time, subjects naturally fixated on the targets at different times. To accommodate this expected variability, we labeled every sample in the epoch with its corresponding BCI score to improve co-registering of target-detections across the group. cBCI scores were computed frame-by-frame by averaging any BCI scores that existed in that frame for a given subset of the subjects. We then compared those values to ground truth labels. Ground truth 'target' labels were generated for every frame that a target item was on-screen.

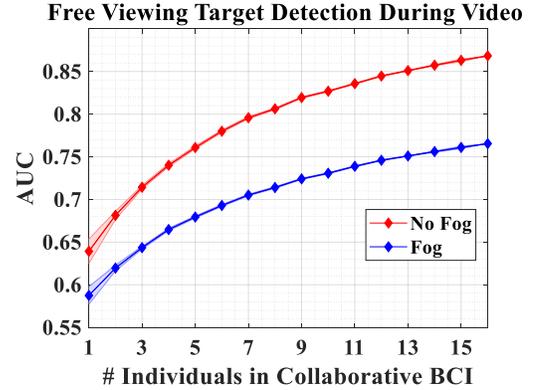

Fig. 3. cBCI group size effect with std. error. Each point is an average of at most 500 unique subject combinations.

We investigate the effect of cBCI group size on the Area Under the Curve (AUC) metric. For statistical testing purposes, we sample, using a bootstrap permutation procedure, a maximum of 500 unique subject combinations per group size, since the number of unique combinations grows combinatorially. We then test the effect of adding additional subjects to the cBCI by using unpaired t-tests with a p-value correction for multiple comparisons using the False Discovery Rate (FDR) procedure by [31]. We also describe the magnitude and time course of the cBCI scores across No-Fog and Fog conditions, along with summaries of associated neural and behavioral responses in the test dataset.

## IV. RESULTS

Both conditions show improved performance with the addition of more subjects into the cBCI ensemble. For both conditions, the only points that were not significant (p>0.05) were changes from 14-15, and 15-16 subjects. The AUCs for the cBCI with all 16 subjects are: 0.8683 for the Non-Fog condition, and 0.7655 for the Fog condition.

Visually inspecting the full cBCI in Fig. 4, the classifier scores appear higher during the video frames when target items appear on-screen. Table II expands on the cBCI peak value and peak times during video frames with target ground truth labels. The peak cBCI score is higher, and peak time is sooner, in the No-Fog condition (p<0.01). Figure 5 illustrates the average time course of the cBCI scores, time-locked to target item onset, for both Fog and No-Fog conditions.

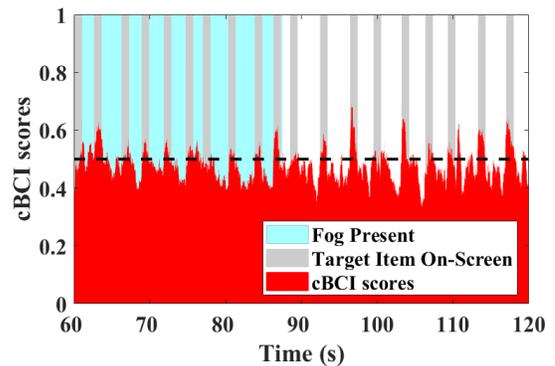

Fig. 4. Co-registered Target Item occurrence and full cBCI classifier scores.

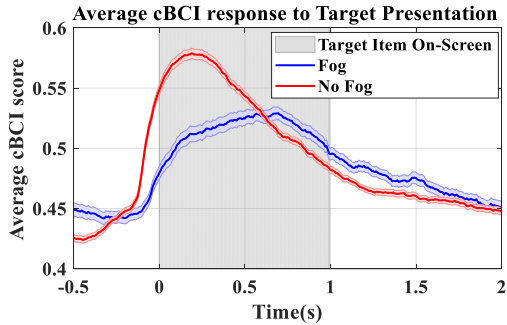

Fig. 5. Comparison of cBCI response to Target Item onset for Fog and No-Fog Conditions with std. error.

The grand-averaged ERPs across all trials, and subjects (electrode Pz) are shown in Fig. 6 for each condition. Table III summarizes behavioral events and timing in the test dataset.

## V. DISCUSSION AND CONCLUSION

In this work we investigate the utility of cBCI in a naturalistic free-viewing environment. We test an experiment-agnostic approach, necessary for real-world scenarios where no training data matching the exact task or circumstances will exist, on a visual target detection task where subjects watch a video under high (No-Fog) and low (Fog) visibility conditions.

We see significantly improved AUC, in most cases, with the addition of one or more subjects into the ensemble. The AUC values appear to be approaching, but did not reach, ceiling performance. The performance may improve further with more subjects. Additionally, we illustrate (Fig.4) that the timing of target items within a video can be reconstructed using cBCI, indicating that such a system has the ability to intelligently filter complex environments based on human-interest.

With this test dataset, we expect the Fog condition to represent a more realistic visual search scenario where targets do not "pop" into and out of existence. Although items pop into existence in in both conditions, the Fog heavily obscures the scene, and likely reduces pop-up effect. As such, we expect subjects to search for target stimuli in the Fog.

This is supported by the results in a few ways: first, the AUCs are lower in the Fog condition. In Fig. 4 and Fig 5. it seems that the Fog-cBCI scores tend to be smaller in magnitude, and rise and fall at a slower rate, than the No-Fog-cBCI scores. However in Table II, we notice that the average peak amplitudes, while significantly different from each other, are closer in value (Fog: 0.5810, No-Fog: 0.6088) than Fig. 5 suggests. Additionally, the average peak time for Fog happens approximately 0.35s later, and has a standard error almost twofold greater, than in the No-Fog condition. This suggests greater variability and latencies in the timing of target-related responses relative to target item onset in the Fog condition. This is corroborated by the grand-averaged ERPs in Fig. 6.

TABLE II. PEAK VALUES AND PEAK TIMES FOR CBCI SCORES TIME-LOCKED TO TARGET ITEM ONSET (MEAN AND STD. ERROR)

| Condition | Average Peak cBCI score | Average Peak Time (s) |
|---|---|---|
| Fog | 0.5810 (±0.0045) | 0.7463 (±0.0556) |
| No-Fog | 0.6088 (±0.0034) | 0.3998 (±0.0315) |

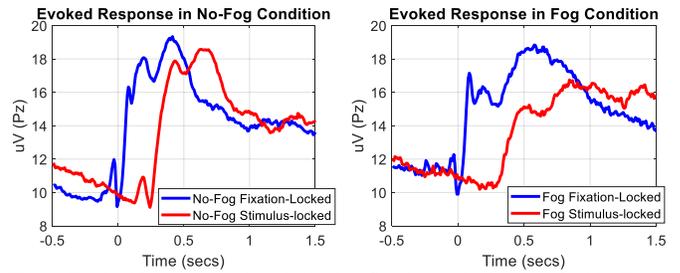

Fig. 6. Evoked Responses, time-locked to fixation and stimulus, for No-Fog (left) and Fog (right) conditions of dataset.

For the No-Fog, the fixation- and stimulus-locked ERPs are almost identical except for a time shift, and the fixation-locked Fog-ERP is relatively similar in shape to the No-Fog ERPs. The stimulus-locked Fog-ERP has a very different profile than the others, implying that this response is less of a function of stimulus onset time than the No-Fog response. Furthermore, in Table III, we see that the time to fixation for the Fog condition has a higher standard deviation than in No-Fog.

Characterizing the Fog and No-Fog conditions provides insight into how to design cBCI systems for dynamic environments. cBCIs should operate robustly in the face of delayed and variable fixations, a feature which distinguishes the Fog condition from the No-Fog. In Fig. 4 and Fig. 5, the higher cBCI scores are not uniformly higher during the times when a target item is present on screen. It is possible that there are better ways to coregister BCI scores and assign ground truth labels to the video frames. Future work will investigate this. However, these ground truth labels will be largely dependent on the eventual applications. For example, higher scores for only a small duration of a target occurrence may be sufficient to detect the item, if that is the only goal. Although out of the scope of this paper, there is a need to better define a "hit" or a "miss" in the context detecting a relevant item in a video or other dynamic environment.

This work demonstrates the utility of cBCI in a dynamic and complex environment. This approach was tested on stimuli more reminiscent of constrained laboratory experiments (No-Fog), and those which more closely approximated real world visual search (Fog). The tested BCI paradigm is also fixation-locked; it does not depend on knowledge of the stimulus timing in order to make a BCI prediction. Additionally, this cBCI approach works with a model that was trained on no data from the test experiment. Thus, the model generalized completely from a pool of multiple, similar, experiments. This indicates the applicability of this approach as a means to leverage expert consensus, via BCI, for reconstructing interesting object occurrence in unseen scenarios.

TABLE III. TIMING OF BEHAVIOR EVENTS IN TEST SET (MEAN AND STD.)

| Condition | Reaction Time Stimulus Locked | Time to Fixation Stimulus Locked | Reaction Time Fixation Locked |
|---|---|---|---|
| Fog | 0.95(±0.23) | 0.45(±0.23) | 0.49(±0.18) |
| No-Fog | 0.71(±0.17) | 0.31(±0.15) | 0.41(±0.13) |

ACKNOWLEDGMENT

This research was sponsored by the Army Research Laboratory under ARL-74A-HRCYB and through Cooperative Agreement Number W911NF-10-2-0022. The views and conclusions contained in this document are those of the authors and should not be interpreted as representing official policies, either expressed or implied, of the Army Research Laboratory or the U.S. Government. The U.S. Government is authorized to reproduce and distribute reprints for Government purposes notwithstanding any copyright notation herein.

REFERENCES

[1] B.P. Bailey and J. A. Konstan, "On the need for attention-aware systems: Measuring effects of interruption on task performance, error rate, and affective state," Computers in Human Behavior, vol. 22, no. 4, pp. 685–708, 2006.

[2] J. M. Puncochar and P. W. Fox, "Confidence in Individual and Group Decision Making: When 'Two Heads' Are Worse Than One.," Journal of Educational Psychology, vol. 96, no. 3, pp. 582–591, 2004.

[3] C. Pavitt, "Colloquy: Do Interacting Groups Perform Better Than Aggregates of Individuals? Why We Have to Be Reductionists About Group Memory," Human Communication Research, vol. 29, no. 4, pp. 592–599, Jan. 2003.

[4] J. R. Wolpaw, N. Birbaumer, D. J. Mcfarland, G. Pfurtscheller, and T. M. Vaughan, "Brain–computer interfaces for communication and control," Clinical Neurophysiology, vol. 113, no. 6, pp. 767–791, 2002.

[5] P. Sajda, E. Pohlmeyer, J. Wang, L. C. Parra, C. Christoforou, J. Dmochowski, B. Hanna, C. Bahlmann, M. K. Singh, S. F. Chang, "In a blink of an eye and a switch of a transistor: cortically coupled computer vision," Proc. IEEE, vol. 98, no. 3, pp. 462-478, Mar 2010.

[6] Polich, J., 2007. Updating P300: an integrative theory of P3a and P3b. Clinical neurophysiology, 118(10), pp.2128-2148.

[7] D. M. Twomey, P. R. Murphy, S. P. Kelly, R. G. O'Connell, "The classic P300 encodes a build-to-threshold decision variable," Eur. J. Neurosci., vol. 42, no. 1, pp. 1636-43, Jul 2015.

[8] F. Kirschner, F. Paas, and P. A. Kirschner, "A Cognitive Load Approach to Collaborative Learning: United Brains for Complex Tasks," Educational Psychology Review, vol. 21, no. 1, pp. 31–42, Dec. 2008.

[9] D. Valeriani, R. Poli, and C. Cinel, "A collaborative Brain-Computer Interface for improving group detection of visual targets in complex natural environments," 2015 7th International IEEE/EMBS Conference on Neural Engineering (NER), 2015.

[10] D. Valeriani, C. Cinel, and R. Poli, "A Collaborative BCI Trained to Aid Group Decisions in a Visual Search Task Works Well with Similar Tasks," in The First Biannual Neuroadaptive Technology Conference.

[11] D. Valeriani, C. Cinel, and R. Poli, "Group Augmentation in Realistic Visual-Search Decisions via a Hybrid Brain-Computer Interface," Scientific Reports, vol. 7, no. 1, Oct. 2017.

[12] L. Korczowski, M. Congedo, and C. Jutten, "Single-trial classification of multi-user P300-based Brain-Computer Interface using riemannian geometry," 2015 37th Annual International Conference of the IEEE Engineering in Medicine and Biology Society (EMBC), 2015.

[13] P. Yuan, Y. Wang, W. Wu, H. Xu, X. Gao, and S. Gao, "Study on an online collaborative BCI to accelerate response to visual targets," 2012 Annual International Conference of the IEEE Engineering in Medicine and Biology Society, 2012.

[14] Y. Wang, Y.-T. Wang, T.-P. Jung, X. Gao, and S. Gao, "A collaborative brain-computer interface," 2011 4th International Conference on Biomedical Engineering and Informatics (BMEI), 2011.

[15] R. Poli, D. Valeriani, and C. Cinel, "Collaborative Brain-Computer Interface for Aiding Decision-Making," PLoS ONE, vol. 9, no. 7, 2014.

[16] H. Touyama, "A collaborative BCI system based on P300 signals as a new tool for life log indexing," 2014 IEEE International Conference on Systems, Man, and Cybernetics (SMC), 2014.

[17] B. O. Mainsah, L. M. Collins, K. A. Colwell, E. W. Sellers, D. B. Ryan, K. Caves, and C. S. Throckmorton, "Increasing BCI communication rates with dynamic stopping towards more practical use: an ALS study," Journal of Neural Engineering, vol. 12, no. 1, p. 016013, 2015.

[18] B. Rivet, A. Souloumiac, V. Attina, and G. Gibert, "xDAWN Algorithm to Enhance Evoked Potentials: Application to Brain–Computer Interface," IEEE Transactions on Biomedical Engineering, vol. 56, no. 8, pp. 2035–2043, 2009.

[19] A. R. Marathe, A. J. Ries, and K. Mcdowell, "Sliding HDCA: Single-Trial EEG Classification to Overcome and Quantify Temporal Variability," IEEE Transactions on Neural Systems and Rehabilitation Engineering, vol. 22, no. 2, pp. 201–211, 2014.

[20] D. Valeriani, R. Poli, and C. Cinel, "A collaborative Brain-Computer Interface to improve human performance in a visual search task," 2015 7th International IEEE/EMBS Conference on Neural Engineering (NER), 2015.

[21] A. Azizian, A. L. Freitas, T. D. Watson, N. K. Squires, "Electrophysiological correlates of categorization: P300 amplitude as index of target similarity," Biol. Psychol., vol. 71, no. 3, pp. 278–288, 2006.

[22] J.R. McDaniel, S.M. Gordon, A.J. Solon, and V.J.Lawhern, "Analyzing P300 Distractors for Target Reconstruction," 2018 Annual International Conference of the IEEE Engineering in Medicine and Biology Society, 2018.

[23] A. J. Solon, S. M. Gordon, V. J. Lawhern, and B. J. Lance, "A Generalized Deep Learning Framework for Cross-Domain Learning in Brain Computer Interfaces," in The First Biannual Neuroadaptive Technology Conference.

[24] S. M. Gordon, M. Jaswa, A. J. Solon, and V. J. Lawhern, "Real World BCI," Proceedings of the 2017 ACM Workshop on An Application-oriented Approach to BCI out of the laboratory - BCIforReal 17, 2017.

[25] Lawhern, V.J., Solon, A.J., Waytowich, N.R., Gordon, S.M., Hung, C.P. and Lance, B.J., 2016. Eegnet: A compact convolutional network for eeg-based brain-computer interfaces. arXiv preprint arXiv:1611.08024.

[26] Touryan, J., Marathe, A. and Ries, A., 2014. P300 variability during target detection in natural images: Implications for single-trial classification. Journal of Vision, 14(10), pp.195-195.

[27] Ries, A.J., Touryan, J., Ahrens, B. and Connolly, P., 2016. The impact of task demands on fixation-related brain potentials during guided search. PloS one, 11(6), p.e0157260.

[28] H. Cecotti, A. R. Marathe, and A. J. Ries, "Optimization of Single-Trial Detection of Event-Related Potentials Through Artificial Trials," IEEE Transactions on Biomedical Engineering, vol. 62, no. 9, pp. 2170–2176, 2015.

[29] Bradford JC, Lukos JR, Reis AJ, Ferris DP (2016). Effect of locomotor demands on cognitive processing. IEEE Engineering in Medicine and Biology Conference, Orlando, FL: August 2016.

[30] L. N. Kaunitz, J. E. Kamienkowski, A. Varatharajah, M. Sigman, R. Q. Quiroga, and M. J. Ison, "Looking for a face in the crowd: Fixation-related potentials in an eye-movement visual search task," NeuroImage, vol. 89, pp. 297–305, 2014.

[31] Benjamini, Y. and Yekutieli, D. The control of the false discovery rate in multiple testing under dependency. Annals of Statistics. 1165-1188, 2001.